# SEIGuard: An Authentication-simplified and Deceptive Scheme to Protect Server-side Social Engineering Information Against Brute-force Attacks

Zuoguang Wang,[1,2,*] Jiaqian Peng,[1,2] Hongsong Zhu,[1,2,*] and Limin Sun[1,2]

[1] School of Cyber Security, University of Chinese Academy of Sciences, Beijing 100049, China.
[2] Beijing Key Laboratory of IoT Information Security Technology, Institute of Information Engineering, Chinese Academy of Sciences (CAS), Beijing 100093, China.

Correspondence: Zuoguang Wang (wangzuoguang16@mails.ucas.ac.cn) and Hongsong Zhu (zhuhongsong@iie.ac.cn).

## Abstract

This paper proposes an authentication-simplified and deceptive scheme (SEIGuard) to protect server-side social engineering information (SEI) against brute-force attacks. In SEIGuard, the password check in authentication is omitted and this design is further combined with the SEI encryption design using honey encryption. The login password merely serves as a temporary key to encrypt SEI and there is no password plaintext or ciphertext stored in the database. During the login, the server doesn't check the login passwords, correct passwords decrypt ciphertexts to be correct plaintexts; incorrect passwords decrypt ciphertexts to be phony but plausible-looking plaintexts (sampled from the same distribution). And these two situations share the same undifferentiated backend procedures. This scheme eliminates the anchor that both online and offline brute-force attacks depending on. Furthermore, this paper presents four SEIGuard scheme designs and algorithms for 4 typical social engineering information objects (mobile phone number, identification number, email address, personal name), which represent 4 different types of message space, i.e. 1) limited and uniformly distributed, 2) limited, complex and uniformly distributed, 3) unlimited and uniformly distributed, 4) unlimited and non-uniformly distributed message space. Specially, we propose multiple small mapping files strategies, binary search algorithms, two-part HE (DTE) design and incremental mapping files solutions for the applications of SEIGuard scheme. Finally, this paper develops the SEIGuard system based on the proposed schemes, designs and algorithms. Experiment result shows that the SEIGuard scheme can effectively protect server-side SEI against brute-force attacks, and SEIGuard also has an impressive real-time response performance that is better than conventional PBE server scheme and HE encryption/decryption. Besides, SEIGuard is also a general architecture applicable to protect other kinds of server-side information.

## 1. Introduction

As an infamous but quite popular attack, social engineering has posed an increasingly serious threat [1]. It describes a type of attack in which the attacker exploit human vulnerability through social interaction to breach cyber security. In the first attack stage, attackers usually make great efforts to gather various kinds of information to learn the targets and to identify their human vulnerabilities. Subsequently, attackers contact with the targets and launch the attack. Thus, in many scenarios, the success of social engineering attacks relies heavily on the information obtained, such as username, password, real name, mobile phone number, email address, identification number, social security ID and bank card number, and these information are known as social engineering information (SEI) [2].

Attackers obtain social engineering information mainly from two sources, open information on the internet and restricted information on the server. And in general the server-side information is more valuable and credible. The leakage and abuse of social engineering information is not only the critical influence factor of social engineering attacks, but also one of the important reasons that lead to further advanced cyber-attack (target attack, APT attack, etc.) and other cybercrimes (spam emails, internet fraud, etc.). Thus, protecting server-side social engineering information is very crucial and significate.

However, users usually select simple, short or easy to remember passwords to server as account login credentials. Besides, some service providers still store users' password and information by plaintext in the database. Once the password was guessed/cracked or the database was pulled off, attacker directly obtains users' sensitive information. For users and service providers with a higher security awareness and IT resources, the former are likely to employ a relatively complex (e.g. more than 8 characters) and the latter will encrypt the password and then store, e.g. by hash or salting hash function, and even the users' information also might be stored in cipher text. When a user password is submitted, the server will check whether the *hash (password)* match with the hash value of credentials stored in the database. If it does not match, the server responses prompting message and refuse the login request. But if it matches, the server will start another process/program branch, decrypt the cipher text and response, and the user will log in successfully. Overall, both the above user authentication and server-side information process are password based encryption (PBE) scheme (Figure 2 shows a more complete flow). However, these widespread PBE login authentication and data encryption schemes have a fatal weakness, which is also the vulnerability that attackers exploit to crack server-side social engineering information, i.e. they are vulnerable to brute-force attacks.

These existing schemes leave attackers with anchors for both online and offline brute-force attacks. Attackers can conduct continuous online password guessing or cracking based on the prompting message that indicates the attempted password is wrong. If the server/database was compromised or the attacker obtained the encrypted password manager/vault or ciphertext of user information, the offline password guessing or cracking is still possible based on the feature difference between the output of wrong guessing and correct guessing [12]. In general, attackers use weak password dictionary and hash/rainbow tables to conduct the brute-force attacks, yet some special / targeted password dictionary might also be created and employed. Moreover, many brute-force tools are easily available on the internet. For instance, HashCat [3], an open source password recovery tool, can crack any hash ciphertext of eight-character-length password in under 2.5 hours by utilizing eight Nvidia GTX 2080Ti GPUs [4], no matter how complex the combination of available 95 basic ascii characters (i.e. printable / non-control characters) [5] in the password is. This comes not long after the news that 620 million hacked accounts went on sale on the dark web [4]. Some measures were deployed to mitigate the online brute-force cracking, such as account locking if frequent wrong password was checked, CAPTCHA and password-based multi-factor authentication. However, the low frequency yet persistent brute-force attack is still feasible, and the CAPTCHA can be also bypassed or cracked. In addition, due to the difference of process/program branch, the attacker can measure the power/energy trail, and the side channel password brute-force cracking is feasible.

Juels and Ristenpart [6] proposed honey encryption (HE) scheme to provide resilience against brute-force attacks. When the ciphertext is decrypted with wrong keys, instead of generating some error or odd output, the decryption will emit plausible-looking but bogus plaintexts called honey messages [6]. HE provides security that never worse than existing PBE schemes, and it benefit especially in cases where too little entropy is available to withstand brute-force attacks that try every key [6]. Afterwards, HE was used by some studies to protect password vault [7], genomic data [8], credit cards No. [9], mobile phone No., identification No. and debit card passwords [10]. However, these studies mainly focus on the data encryption (confidentiality), the problem about server-side response and login authentication were not concerned. However,

HE is mainly suitable for information object with simple or fixed format (e.g. 6-digit password [10], simple numeric value [16]), the message space of which are usually small and limited, and the message distribution is uncomplicated. Honey encryption and decryption involve a large amount of computing, read operation and write operation. The (computing, RAM, storage) overhead of processing a large message space is very high [10]. For a $10^8$ message space (8-digit number), the decryption time will reach 180 seconds [10], let alone for phone number, identification No. and credit/debit card No. Mok et al. [11] attempt to use HE process file name, but "the message space of the proposed scheme is pre-fixed and the complexity and the size of inverse sampling tables is growing exponentially with the increase of the file names and its extension sizes. [11]" Therefore, for server application scenario, the design and algorithm of HE scheme for specific information object and the real-time performance of server response are two difficult problems. Studies [13] and [14] combined honey encryption and honeyword to strengthen the password storage, and alert the attack tries if the entered password is checked to be invalid by original password stored in files or database. Tan et al. [15] [16] also discussed an extended HE scheme to enhance password-based authentication, in which the password are processed by HE, and if the submitted password is checked to be invalid the user will be redirect to a fake account and triggers an alert. Study [16] also showed HE application case on diabetes data (simple numeric value). These studies involved or strengthened online user authentication, but HE application on other slightly complex information objects and server response performance were not concerned. What's more important is that these scheme are still based on PBE authentication, and the anchor that brute-force attacks relying on still remains. In other words, the (online, offline or side channel) brute-force attacks is still possible.

To summarize, there are still many challenges to protect social engineering information, especially for complex server-side social engineering information against brute-force attacks.

This paper makes the following contributions.

- This paper proposes an authentication-simplified and deceptive protection scheme (SEIGuard) for server-side social engineering information. In this scheme, the password based encryption (PBE) *login check and branches* were removed to eliminate the anchor point that the online and offline brute-force attacks depending on, and the login password merely serves as a temporary key to generate the cipher text by honey encryption and is later dropped. The honey plaintext obtained by incorrect login password will mislead attackers, and the scheme can resist the online brute-force attacks, and even the offline attack where the attacker get the server-side database. No matter the inputted passwords are correct or wrong, they share the same undifferentiated process, which depresses the side channel password attack.
- This paper proposes the designs and algorithms of HE (distribution transforming en/decode, honey en/decryption) for not only limited & uniformly distributed message space (e.g. phone number, ID) but also for unlimited & uniformly distributed message space (e.g. email address) and unlimited & non-uniformly distributed message space (e.g. personal real name). These design involve processes such as information registration, information storage, user login authentication and information response. The algorithms are low in time complexity and efficient in real-time response.
- This paper implement the SEIGuard system based on the proposed schemes, designs and algorithms. The experiment evaluation shows that the SEIGuard scheme is effective in protecting server-side (social engineering) information against brute-force attacks, and the real-time response performance is also very impressive.

The reminder of this paper is organized as follows. Section 2 describes the related work. Section 3 presents the SEIGuard scheme, as well as specific schemes, designs and algorithms for different kinds of information object (and message space). Section 4 conducts contrast experiments to evaluate the security and real-time response performance of SEIGuard. Section 5 is discussion. Section 6 concludes the paper.

## 2. Related Work

Honey Encryption [6] consist a pair of algorithms, HE = (HEnc, HDec), as Figure 1 shows. Encryption algorithm HEnc takes a key $K \in K$, message $M \in M$, possible random bits as input, and outputs a ciphertext C. This can be written as C ← $ HEnc (K, M), where ← $ denotes that HEnc may use some number of uniform random bits. Decryption algorithm HDec takes a key K ∈ $K$, ciphertext C as input, and outputs a message M ∈ $M$. Decryption is deterministic and can be written as M ← HDec (K, C).

| Honey Encryption Algorithm | Honey Decryption Algorithm |
| --- | --- |
| $ Henc (K, M) | Hdec (C, K) |
| S ← $ DTE-encode (M) | S ← decrypt (C, K) |
| C ← $ encrypt (K, S) | M ← DTE-decode (S) |
| return C | return M |

Figure 1: Honey encryption and decryption algorithm

HE has a two-step procedure called DTE-then-encrypt, where the distribution-transforming encoder (DTE) is a randomized message encoding algorithm that served as the cornerstone of HE scheme.

First, the DTE was designed with an estimate of the message distribution *pm*, and the message distribution was transformed/mapped to seed that distributed (approximately) uniformly. Encoding a message M sampled from *pm* by DTE will yield a seed value S ∈ $S$. Then, the seed S is encrypted by a conventional encryption algorithm *encrypt()* using the key K, and the output is the HE ciphertext C. On the other hand, when decrypting the ciphertext C, the seed S was firstly generated by the corresponding conventional decryption algorithm *decrypt()* with K. Then the message M is decoded by the reverse process of DTE.

If an incorrect key K' is used to crack the ciphertext M, a valid but incorrect seed S' ∈ $S$ will be firstly generated. Subsequently, instead of arising errors or output odd strings, a honey message M' ∈ $M$ (that subjects to the same *pm* and looks plausible with M) will be decode by DTE and returned to the attacker.

Thus, HE, more specifically DTE(encode, decode), provides the ciphertext an additional protection and deception layer. The attackers are unable to succeed in message recovery even after trying every possible key [6]. If the designed DTE is only approximately fit the message distribution *pm*, HE can nevertheless prove message-recovery security far beyond the brute-force-barrier. If the DTE is bad, i.e., based on a poor estimate of *pm*, HE will fall back to normal security. In other words, HE never do worse than PBE schemes [6].

## 3. SEIGuard Scheme and System

In this section, we first intrude the overall scheme (SEIGuard) that proposed to protect server-side social engineering information (SEI) against brute-force attacks. It is an authentication-simplified and deceptive scheme using HE, which contains special designed procedures such as information registration (honey encryption), data storage, login authentication and (honey decryption) server response. Then we respectively describe the scheme design and algorithm for 4 typical social engineering information objects, which represent 4 different types of message space. They are different designs under the same architecture/scheme (SEIGuard), since different message space needs customized HE design and implementation. Finally, we develop the SEIGuard system, in which all of the above schemes are integrated.

## 3.1. SEIGuard: an authentication-simplified and deceptive scheme to protect server-side data against brute-force attacks

Figure 2 describes the current password-based (user and information) register, login authentication and server response flows.

- In the account register stage, a user submits username, password and other information. If the username is available, the server will process the account details and then store the corresponding plaintext or ciphertext in the database. In general, the user password will store in plaintext, hash or salting hash (as key).
- When a user attempt to log in the system using certain username and password, the server will first process and check the login credentials (username and password) where the key authentication is that for a specific username if the processed password is not match with the key stored in the database, the authentication will be treated as failed, and the server then will response prompting message like "the password is wrong, please reinput"; otherwise, if the processed password is match with the key stored in the database, the server then will process and response plaintext information to the user.

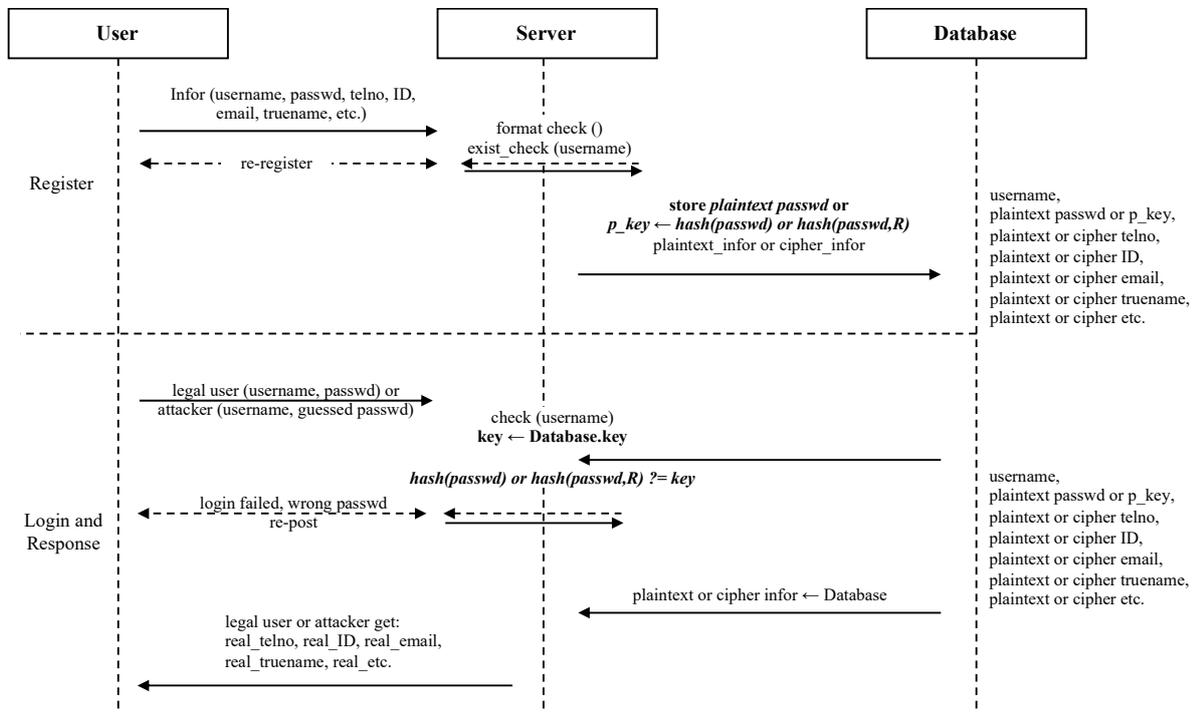

Figure 2: Current password-based register, login authentication and server response flows

As described in Introduction, this scheme is vulnerable to brute-force attacks. It leave attackers with anchors for both online and offline brute-force attacks. Attackers can conduct persistent online password guessing or cracking based on the prompting message that indicates the attempted password is wrong. If the server/database was compromised or the attacker obtained the ciphertext, the offline password cracking is still possible based on the feature difference between the output of wrong guessing and correct guessing [12]. In addition, due to the difference of process/program branch, the attacker can measure the power/energy trail, and the side channel password brute-force cracking is feasible.

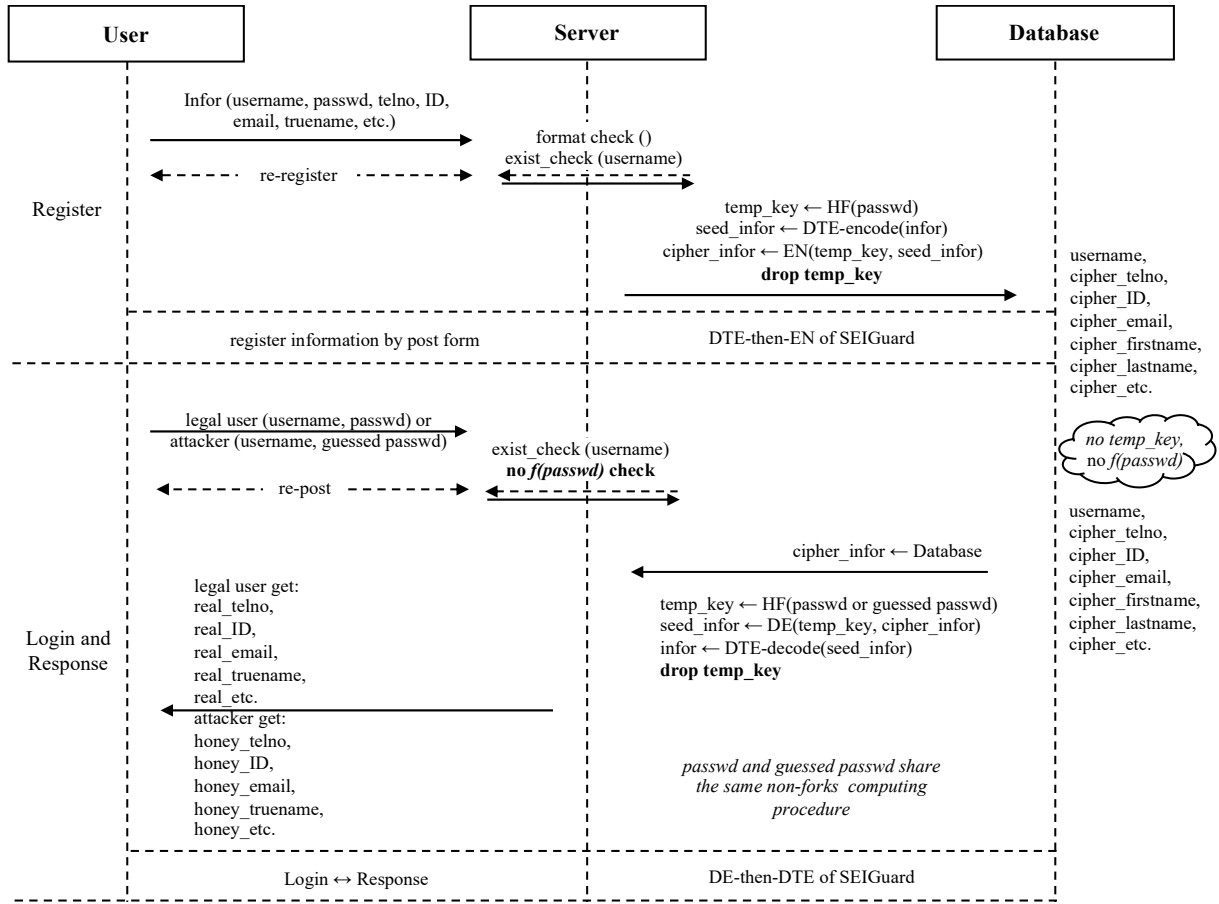

Figure 3: register, login authentication and response flow of SEIGuard scheme

This paper propose an authentication-simplified and deceptive scheme to protect server-side social engineering information against brute-force attacks, as Figure 3 shows. It is different from the current / conventional password-based authentication and data protection scheme.

- In the account register stage, a user submits username, password and other information. If the username is available, the server first process the password to a temporary key (e.g. by a certain hash function, md5), and then other user information will be encrypted by honey encryption (HE) using the temporary key *tmp_key*. The information is firstly randomly encoded to a seed, and then the seed is encrypted to ciphertext by algorithm EN using *tmp_key*. What's ingenious is that subsequently the temporary key is dropped and other ciphertexts are stored in the database, rather than storing them all.
- When a user attempt to log in the system using certain username and password, the server merely checks whether the username is valid (this is optional) and doesn't check the password.
  - ◇ If the username is valid (e.g. existed), the server will first process the password to a temporary key (e.g. by the same hash function) and then directly decrypt the user's ciphertexts (generated by previous HE) by DE using this temporary key. If the EN is reversible, EN = DE. By virtue of the later designed honey decryption scheme, there will always be a valid seed generated and a plausible plaintext returned, no matter the submitted password is correct or wrong. And the correct password will yield the original/real/correct plaintext but the correct password yield honey plaintext.

✧ If the username is invalid, "the username or password is wrong, please resubmit" will also serve as a deceptive prompting message responded to the user / attacker.

In the proposed scheme, the password check is intentionally omitted and there is no password or encrypted/processed password stored in the database, and this design is further combined with the data encryption design using honey encryption. This scheme not only eliminate the anchor in authentication that brute-force attacks relies on, but also remove the cracking object (ciphertext of password). Even the ciphertexts of other different information objects are indistinguishable, since they share the same temporary key and the same ciphertext space (Figure 4). Besides, every guessing/brute-force attacks share all the same undifferentiated procedures from login to response, no matter the password is correct or wrong. And no visual difference and statistic difference can be found from the outputs of brute-force attacks, since all the (correct and honey) outputs obey the same (original) message (probability) distribution. Therefore, the anchors about data decryption that brute-force attacks relies on are further removed. Consequently, even if the server is compromised, i.e. the attacker have access to database, mapping files and server-side source code, the offline brute-force attacks are still impossible (or extremely hard). Overall, this is an authentication-simplified and deceptive scheme preventing both online, offline and side-channel brute-force attacks.

Next 4 subsections, we respectively describe the specific scheme design and algorithm for 4 typical social engineering information objects (mobile phone number, identification number, email address, personal name), which represent 4 different types of message space.

For each case, there are 3 core stages and procedures need careful designs in SEIGuard scheme: create message mapping files (for DTE encode and decode), register information (honey encryption and ciphertext storage), and response login (user check and honey decryption).

## 3.2. Scheme design and algorithm for limited message space with uniform distribution: information objects such as mobile phone number, pin code and password

Mobile phone number is a typical social engineering information (SEI) that attackers contact with the targets and launch social engineering attacks such as pretexting, vishing and smishing. Besides, it is also important SEI that attackers learn and investigate the target, since more and more accounts are bound to mobile phone number, and many online activities can be correlated by it.

Although mobile phone number in different countries are various, they share the similar composition. It has a certain fixed length and correspondingly a limited message space, in which the numbers obey uniform distribution. In Chinese, the mobile phone number consists of 11 digits. The first 3 digits represent the operator code, the 4th to 7th digits are the location code and the 8th to 11th digits are random [10]. Therefore, scheme designs are the same and the algorithms for other countries' phone number are analogous.

Algorithm 1.1 describes the procedure to create message mapping files for mobile phone number, which is the basis for subsequent seed sampling in encryption and reverse sampling in decryption. According to [10], for a $10^8$ (8 digits) message space, the encryption and decryption time will reach 70s and 180s, let alone for phone number (11 digits) message space which is nearly $10^{11}$. This bring about a big challenge, since it poses a serious negative effect on the real-time response of online service. To address this problem, we employ a multi-files strategy for mapping files, i.e. we divide the mapping files into multiple small files, and design a binary search algorithm in Algorithm 1.2, to boost the time performance of honey encryption and decryption. Every small mapping file was designed to contain 100000 lines based on our experiments. Besides, instead of storing all the mapping details from phone No. to every piece of seed, we merely store the cumulative probability (CP) value of the last seed mapped to every phone No.. By this way, we calculate the seed sampling (Algorithm 1.2) in encryption rather than search in the mapping file.

---

| Algorithm 1.1: distribution transforming encode of mobile phone number in SEIGuard |
| --- |

**Input:** mobile phone No. message space

**Output:** mapping files of phone No. message space

---

1:   **procedure** create_phone_no_mapfile(m_sum, f_unit)          $\triangleright$ f_unit = 100000

2:       prob_density $\leftarrow$ float(1) / m_sum      $\triangleright$ probability density of uniform distribution

3:       f1 $\leftarrow$ open("telno.txt", 'r')

4:       lines $\leftarrow$ f1.readlines("telno.txt")

5:       file_count $\leftarrow$ int(m_sum / f_unit) + 1      $\triangleright$ multiple small mapping files strategy

6:       j $\leftarrow$ 0

7:       **for** i in range(file_count) **do**

8:          f2 $\leftarrow$ open("phone_no_file" + str(i) + ".txt", 'w')

9:          **if** i $\neq$ file_count $-$ 1 **then**

10:            lines_slices $\leftarrow$ lines[i * f_unit: (i + 1) * f_unit]

11:          **else**

12:            lines_slices $\leftarrow$ lines[i * f_unit:]

13:          **end if**

14:          **for** line in lines_slices **do**

15:            j $\leftarrow$ j + 1

16:            lineinf2 $\leftarrow$ line.strip() + ',' + str(j * prob_density)   $\triangleright$ form: phone no. , seed CP

17:            f2.write(lineinf2)

18:          **end for**

19:       **end for**

20:       f1.close(), f2.close()

21:  **end procedure**

---

Algorithm 1.2 describes the register design of mobile phone number in SEIGuard. To enhance the security, we set the seed space to 2^64 , which is a far larger than the phone number sapce. The user password is first hashed by md5() and processed to a 64 bits temporary key. Then, we calculate the seed block mapped to the user submitted phone number, and randomly sample a seed. This seed is later encrypted by XOR (other encryption is also practicable) using the temporary key, and the generated ciphertext is then stored in the database.

| Algorithm 1.2: register (honey encryption and ciphertext storage) of mobile phone number in SEIGuard |
|---|

**Input:** user submitted mobile phone No.

**Output:** cipher text of phone No. in the database

| | |
|---|---|
| 1: | **procedure** register_HE(username, passwd, phone_no, seed_space)    ▷ seed_space = 2^64 |
| 2: |    **if** not exist_check(username) **then** |
| 3: |       tmp_key ← p_md5(passwd)                    ▷ tmp_key is a 64 bits md5(passwd) |
| 4: |       end ← cumul_distr(phone_no) * seed_space |
| 5: |       start ← (end - prob_density * seed_space) + 1 |
| 6: |       seed ← random(start, end)                        ▷ seed random sampling |
| 7: |       cipher_phone_no ← tmp_key ⊕ seed                 ▷ reversible encryption XOR |
| 8: |       store_database(cipher_phone_no) |
| 9: |    **else** |
| 10: |       prompt("the username has existed") |
| 11: |       **return** false |
| 12: |    **end if** |
| 13: | **end procedure** |

Algorithm 1.3 describes the server response design in SEIGuard. When a user/attacker attempt to log in the system using certain username and password, if the username is valid, the server start the honey decryption. The password is first processed to a 64 bits temporary key, by which the phone No. ciphertext is decrypted and a seed is generated. Then, the DTE decode (reverse sampling) starts. Instead of open a big mapping file and search the corresponding phone No. line by line, we employ a binary search to locate the position/rank of the phone No. in the whole mapping space and then calculate which file and which line the plaintext locates in our small mapping files. Finally, we directly fetch the plaintext in specific file and line, and then response it to the user/attacker. On the other hand, if the username is invalid, the server will response "the username or password is wrong" which is also a deceptive message.

| Algorithm 1.3: login and response (honey decryption) of mobile phone number in SEIGuard |
|---|

**Input:** username, passwd

**Output:** plaintext of phone no in user interface

| | |
|---|---|
| 1: | **procedure** login_HD(username, passwd) |
| 2: |    **if** exist_check(username) **then** |
| 3: |       tmp_key ← p_md5(passwd) |
| 4: |       seed ← tmp_key ⊕ cipher_phone_no |
| 5: |       position ← binary_search(seed)            ▷ locate No. position by  binary search |
| 6: |       file_name ← position / f_unit |
| 7: |       line_no ← position % f_unit +1 |
| 8: |       plaintext ← getline(file_name, line_no)[0]      ▷ fetch plaintext in specific file and line |
| 9: |       **return and response** plaintext |
| 10: |    **else** |
| 11: |       prompt("the username or password is wrong") |
| 12: |       **return** false |
| 13: |    **end if** |
| 14: | **end procedure** |

These design and algorithm are also fully applicable to social engineering information such as pin code (4 - 8 digits, 10^8), bank card password (6 digits, 10^6).

### 3.3. Scheme design and algorithm for limited message space with uniform distribution: complex information objects such as Chinese identification number

Chinese identification number (18 digits) / social security number (9 digits) and bank debit / credit card number (12 -19 digits) are well known confidential information (and also social engineering information). They are in general, they have a larger message space than information objects in section 3.2, and a more complex composition format, e.g. checksum. Therefore, their scheme design and algorithm are also relatively complex.

Chinese identification number consists of 18 digits, in which the last digit is checksum and the other 17 digits are similar to the mobile phone number. Thus, the scheme design and algorithm of creating mapping files are similar as Algorithm 1.1, the same multiple small mapping files strategy is used. What's different is that here we only map the first 17 digits, and the last digit will be cut off in the register procedure and calculated & appended in the response procedure. Hence, we skip the algorithm description (Algorithm 2.1) of creating mapping files (distribution transforming encode) for Chinese identification number.

Algorithm 2.2 describes the register design of identification number in SEIGuard. We add the validity check of ID No. after the check of username. Subsequently, the last digit of the submitted ID No. is dropped and the first 17 digits are processed by honey encryption procedure. The design of seed space ($2^{64}$), file unit (100000), temporary key process, random sampling and encryption function are the same as Algorithm 1.2. We will not repeat them here.

| Algorithm 2.2: register and honey encryption of ID number in SEIGuard |
|---|
| **Input:** user post ID No. |
| **Output:** cipher text of ID No. in the database |
| 1:  **procedure** register_HE(username, passwd, id_no, seed_space) |
| 2:      **if** not exist_check(username) **then** |
| 3:          **if** valid_check(ID_no) **then** |
| 4:              pid_no ← id_no[0:17]) |
| 5:              tmp_key ← p_md5(passwd) |
| 6:              end ← cumul_distr(pid) * seed_space |
| 7:              start ← (end - prob_density * seed_space) + 1 |
| 8:              seed ← random(start, end) |
| 9:              cipher_pid_no ← tmp_key ⊕ seed |
| 10:             store_database(cipher_pid_no) |
| 11:         **else** |
| 12:             prompt("the ID N.O. is not valid") |
| 13:             return false |
| 14:         **end if** |
| 15:     **else** |
| 16:         prompt("the username has existed") |
| 17:         **return** false |
| 18:     **end if** |
| 19: **end procedure** |

Algorithm 2.3 describes the server response design of Chinese ID number in SEIGuard. The principal part is the same as Algorithm 1.3. The difference is that the output of the honey decryption in this algorithm is the first 17 digits, based on which the last digit is then calculated and the complete ID No. is generated.

| Algorithm 2.3: login and response (honey decryption) of ID number in SEIGuard |
| --- |
| **Input:** username, passwd |
| **Output:** plaintext of ID No. in user interface |
| 1:   **procedure** login_HD(username, passwd) |
| 2:      **if** exist_check(username) **then** |
| 3:         tmp_key ← p_md5(passwd) |
| 4:         seed ← tmp_key ⊕ cipher_id_no |
| 5:         file_name, line_no ← binary_search(seed) |
| 6:         pid ← getline(file_name, line_no)[0] |
| 7:         last_digit ←  calculate_last_digit(pid)    ▷ calculate the last digit/checksum of ID No. |
| 8:         plaintext ← str(pid) + str(last_digit)        ▷ generate the complete ID No. |
| 9:         **return and response** plaintext |
| 10:     **else** |
| 11:         prompt("the username or password is wrong") |
| 12:         **return** false |
| 13:     **end if** |
| 14:  **end procedure** |

These design and algorithm are also fully applicable to information objects like Chinese ID No., e.g. bank debit / credit card number.

### 3.4. Scheme design and algorithm for unlimited message space with uniform distribution: information object like email address

Email address is another typical social engineering information. It is the essential information for attackers to conduct phishing attack and it also sometimes indicate the target's affiliation. According to RFC 5321 [17], the format of an email address is "local-part@domain", in which the maximum length of local-part and domain is respectively 64 and 255 octets. As a result, without considering the domain part, merely the message space of local-part reaches $64^{63}$(alphanumeric characters and ".") ~ $64^{95}$(printable characters). It is too large to generate the whole message space and create the complete mapping files. However, one of the preconditions of the honey encryption scheme is creating a complete message distribution transformation (i.e. creating a complete mapping file). Even if we have the unrestrained storage space, in the context of server response, a bad time performance seems predictable. To cope with this challenge, we treat email address as an unlimited message space, and propose an incremental DTE (and mapping files creation) design/solution.

Algorithm 3.(1-2) describe the register and DTE design of identification number in SEIGuard. In this scheme design, the information registration and the mapping files creation is simultaneous. Due to the local-part of the email address is often used as certain forms of username, the usual email address is not odd-looking. Therefore, we combine the user submitted email address and leaked dataset of real email address to make up the email address (honey) message space, and this space is regarded to obey uniform distribution based on the consideration that all the email addresses have the same or similar real-looking degree (i.e. share an equal probability density). Further, we set the "seed density" to *seed_unit*, i.e. every message (email address) is mapped /

DTEed to *seed_unit (e.g. 10)* seeds, and to enhance the security, every time a new user message registered, another (*seed_block -1, e.g. 9*) honey messages will be randomly sampled from the honey dataset and combined with the user messages in a randomly order. In other words, for every original user message, they are mapped to *seed_unit * seed_block (i.e. 100)* seeds where (*seed_block -1)/seed_block (i.e. 90%)* seeds are honey seed. And the mapping files are increased and created automatically.

When a new user start a register, the server first add {*seed_block* message, *seed_unit * seed_block* seeds} to the mapping files, and then a seed of the user message is randomly sampled and used in the honey encryption. Besides, after increasing the mapping files, we update the seed number that has been assigned, for the future registering. It should be noted that, we prepared a small mapping file in advance, in which a certain amount of (e.g. 10000) email address mapping have existed. This measure is for the purpose of preventing offline brute-force attacks, since the mapping information & files is incremental and it seems a low security if the mapping file has a small data set (when the user registration is at the beginning).

---

Algorithm 3.(1-2): register and    distribution transforming encode of email address in SEIGuard

**Input:** email address message space
**Output:** incremental mapping files of email address message space and ciphertext of email address

1:  **procedure** register_HE(username, passwd, m)
2:      **if** not exist_check(username) **then**
3:          tmp_key ← p_md5(passwd)
4:          seed_num ← db.query.filter_by(name="email").num          ▷ assigned seed number
5:          insert_index ← random.randrange(seed_ block)   ▷ randomize the user email index
6:          increase_mapfiles(m, seed_num, seed_unit, seed_ block, insert_index)          ▷ call 2nd pro
7:          seed ← int(seed_num + (insert_index + random.random())* seed_unit)          ▷ S sampling
8:          cipher_email ← tmp_key ⊕ seed
9:          store_database(cipher_email)
10:     **else**
11:         prompt("the username has existed")
12:         **return** false
13:     **end if**
14: **end procedure**
15: **procedure** increase_mapfiles(m, seed_num, seed_unit, seed_ block, insert_index)
16:     f1 ← open(email-d.txt", "r")                  ▷ email-d.txt is a honey email address dataset
17:     email ← random.sample(f1.readlines(), seed_ block - 1)          ▷ increase email addr space
18:     email.insert(insert_index, m + "\n")                              ▷ insert user email addr
19:     file_sum ← int(seed_num / seed_unit / f_unit)          ▷ multiple small mapping files strategy
20:     f2 ← open("email" + str(file_sum) + ".txt", "a")
21:     **for** i in range(0, seed_ block) **do**                              ▷ create  mapping  files
22:         seed_num ← seed_num + seed_unit
23:         f2.write(email_1[i] + "," + str(seed_num))
24:     **end for**
25:     db.update(seed_num)
26:     f1.close(), f2.close()
27: **end procedure**

Algorithm 3.3 describes the server response design of email address in SEIGuard. The principal part of this design is similar to previous algorithms. What is special is that the seed space is far larger than the message number that have been mapped, and in order to ensure the correctness and validity of the decrypted seed, we employ the modulo operation on it. The correct password will decrypt and modulo a correct seed and then fetch the correct email address, but an incorrect password guessing will decrypt and modulo a honey seed and then fetch a honey email address.

Besides, a parallel system can be used to monitor the status of honey email addresses, which is a measure to detect the possible brute-force attacks.

---

**Algorithm 3.3: login and response (honey decryption) of mail address in SEIGuard**

**Input:** username, passwd

**Output:** plaintext of email address in user interface

| | |
|---|---|
| 1: | **procedure** login_HD_email(username, passwd) |
| 2: |     **if** exist_check(username) **then** |
| 3: |         tmp_key ← p_md5(passwd) |
| 4: |         seed ← tmp_key ⊕ cipher_email |
| 5: |         seed ← seed % seed_num |
| 6: |         file_name, line_no ← binary_search(seed) |
| 7: |         plaintext ← getline(file_name, line_no)[0] |
| 8: |         **return and response** plaintext |
| 9: |     **else** |
| 10: |         prompt("the username or password is wrong") |
| 11: |         **return** false |
| 12: |     **end if** |
| 13: | **end procedure** |

---

The proposed scheme design and algorithms are also applicable to information objects which have similar characteristics as email address.

## 3.5. Scheme design and algorithm for unlimited message space with non-uniform distribution: (Chinese) personal name

Personal (real/true) name is a typical social engineering information, which is the most used identification in a person's social and online life. Various types of information are correlated by a person's name. In situations where a real-name authentication is needed, the person's real name is regarded as confidential information. Even in the common social engineering attack scenarios, knowing the targets' name is a shortcut to obtain their trust.

However, generating a message space that covers all the available personal names seems impracticable. The Chinese characters seem to be countably infinite, the length of a name (even the surname/lastname) is various, and some characters and words have a high frequency in names while some are rarely used. Therefore, the message space of Chinese personal name can be regarded as a complex, unlimited message space with non-uniform distribution.

According to the fact that Chinese personal name has a two-part structure (firstname, lastname) and a valid (name) message space is consist of is consist of a large part of high-frequency characters and a small number of unfamiliar characters, we propose a _two-part HE (DTE), 7 level_ _message distribution and incremental mapping files_ scheme design to process this complex and unlimited message space.

- The _two-part HE (DTE) design_ means that firstname and lastname are processed separately, since they are two different message space. In Chinese personal names, there are 100 familiar surnames, 200 infrequent surnames and some rare surnames. These 300 surnames constitute the main-part of the firstname message space. According to the "the latest population ranking of China's 300 surnames [18][19]", we extract the probability density distribution of these surnames and they are DTEed to the lastname seed space (Algorithm 4.1.2).
- For the firstname, Table of General Standard Chinese Characters (TGSCC) [20] is a very useful material to generate the Chinese personal name message space. This table is studied, created and published by Chinese government based on three main resource: the characters used in Chinese names, place name, science and technology terms and Classical Chinese in teaching material. These characters are categorized into 3 levels. The first level table contains 3500 frequently used characters. The second level table contains 3000 characters that has a relative low frequency compared with the first level. The third level table collects 1650 rarely used characters. Therefore, we set 3 probability weight for these 3 levels characters, i.e. _wl1, wl2, wl3(1, 0.01, 0.0001)_. Besides, considering almost all the Chinese firstname is consist of 1 or 2 characters, we create all the combinations of TGSCC characters, which make up of the main-part of the firstname message space. And as a result, there are 9 probability weight combinations (in 6 different value/level) for the firstname message distribution. Further, we extract 131818 non-repetitive and very familiar firstnames from a Chinese names corpus [21], and set their probability weight to _wl4 (10000)_. Then, the previous tuples (message space, weight) are updated by (very familiar fm, _wl4_). Finally, we get a message distribution that has 7 level probability density, which is approximate the real firstname distribution. (Algorithm 4.1.1)
- The _incremental mapping files design_ (similar as the email scheme design) is also used here, in order to process the rare firstname and lastname. Because they are rare, their probability density are respectively set to the minimum level.

Algorithm 4.1.1 describes the design to generate the main-part message space, DTE and create the mapping files for Chinese firstnames.

| Algorithm 4.1.1: create mapping files of main-part message space of chinese firstname |
| --- |
| **Input:** general standard Chinese characters list (three leves) and familiar firstname list |
| **Output:** main message spaces and mapfiles of chinese firstname |

| | | |
| --- | --- | --- |
| 1: | **procedure** create_firstname_mapfiles(w1.txt, w2.txt, w3.txt, familiar_firstname.txt) | |
| 2: | W ← w1[] + w2[] + w3[] | ▷ 8105 in total |
| 3: | M ← all_combine(w[], len < 3) | ▷ 656 99130 in total |
| 4: | M.sort() | ▷ sort firstname by unicode |
| 5: | weight ← 1, wl1 ← 1, wl2 ← 0.01, wl3 ← 0.0001, wl4 ← 10000 | |
| 6: | **for** m in M **do** | |
| 7: | **for** word in m **do** | |
| 8: | **if** word in w1 **then** | |
| 9: | weight ← weight * wl1 | |
| 10: | **elif** word in w2 **then** | |
| 11: | weight ← weight * wl2 | |

```
12:            elif word in w3 then
13:                weight ← weight * wl3
14:            end if
15:        end for
16:        M2 ← (m, weight)
17:    end for
18:    for fm in familiar_firstname do
19:        if binary_search(fm, M2) then
20:            (fm, weight) ← (fm, wl4)        ▷ for very familiar firstname, update the weight to wl4
21:        end if
22:    end for
23:    seed_unit ← 10,0000,0000, end_seed ← 0
24:    for (m,weight) in M2 do
25:        sblock_size ← weight * seed_unit
26:        end_seed ← end_seed + sblock_size
27:        M3 ← (m, end_seed, sblock_size)
28:    end for
29:    firstname_mapfiles ← divide(M3, file_unit)
30: end procedure
```

Algorithm 4.1.2 describes the procedure of (DTE and) creating the mapping files for the main-part message space of Chinese lastnames.

| Algorithm 4.1.2: create mapping files of main part message space of Chinese surname/lastname |
| --- |
| **Input:** list and frequency of Chinese lastname from official statistics |
| **Output:** main message spaces and mapfile of chinese lastname |

```
1:  procedure create_lastname_maffiles(lastname.txt, seed_initial)
2:      f1 ← open("surname_initial.txt", 'r', encoding='utf-8')
3:      lines ← f1.readlines()
4:      f2 ← open("lastname_mapfile.txt", 'w', encoding='utf-8')
5:      end_seed ← 0
6:      for (m, f) in lines do                    ▷ f is the probability density of m in 300 surnames
7:          sblock_size ← f * seed_initial
8:          end_seed ← end_seed + sblock_size
9:          f2.write(m + "," + str(end_seed) + "," + str(sblock_size))
10:     end for
11:     f1.close(), f2.close()
12: end procedure
```

Algorithm 4.2 describes the register and DTE design of personal name in SEIGuard. Corresponds to the previous design, the firstname and lastname are processed separately using the same procedure, and compared to procedures of previous information object, this procedure is more complex. When a firstname is submitted, the server firstly check whether it exists in the mapping files of the main-part message space (by binary search). If it is existed, randomly sample a seed from the mapped seed block. But if it doesn't exist, the server starts the incremental mapping files design procedure (creates new mapping, randomly samples and increase the mapping files). Finally, the seed is encrypted using the temporary key to be a ciphertext.

---

Algorithm 4.2: register and honey encryption of personal name/truename in SEIGuard

**Input:** username, passwd, truname

**Output:** increased mapping files of firstname and lastname space, and and cipher text

---

1:   **procedure** register_increase_mapfile(username, passwd, fn, ln)
2:       **if** not exist_check(username) **then**
3:           tmp_key ← p_md5(passwd)
4:           f_seed_num ← db.query.filter_by(name="firstname").num
5:           l_seed_num ← db.query.filter_by(name="lastname").num
6:           **if** binarysearch(fn, ln_mapfiles) **then**
7:               f_seed ← end_seed - (random() * sblock_size)            ▷ sampling seed for fn
8:           **else**
9:               f_seed ← int(f_seed_num + random.random() * f_min_seed_unit)
10:              increase_fn_mapfile(fn, f_seed_num, f_min_seed_unit)    ▷ same as algorithm 3.2
11:          **end if**
12:          **if** binarysearch(ln, lastname_mapfiles) **then**
13:              l_seed ← end_seed - (random() * sblock_size)            ▷ sampling seed for ln
14:          **else**
15:              l_seed ← int(l_seed_num + random.random() * l_min_seed_unit)
16:              increase_ln_mapfile(fn, l_seed_num, l_min_seed_unit)    ▷ same as algorithm 3.2
17:          **end if**
18:          cipher_fn ← tmp_key ⊕ f_seed
19:          cipher_ln ← tmp_key ⊕ l_seed
20:          store_database(cipher_fn, cipher_ln)
21:      **else**
22:          prompt("the username has existed")
23:          **return** false
24:      **end if**
25:  **end procedure**

---

Algorithm 4.3 describes the server response design of Chinese personal name in SEIGuard. The firstname and lastname are honey-decrypted separately, and we design different binary search functions for all the involved information locating operations. The procedure of each is similar as the honey decryption of email address, and we will not repeat them here.

| Algorithm 4.3: login and response (honey decryption) of personal name/truname in SEIGuard |
|---|

**Input:** username, passwd

**Output:** plaintext of truname in user interface

```
 1:  procedure login_HD_truname(username, passwd)
 2:      if exist_check(username) then
 3:          tmp_key ← p_md5(passwd)
 4:          f_seed_num ← db.query.filter_by(name="firstname").num
 5:          l_seed_num ← db.query.filter_by(name="lastname").num
 6:          f_seed ← tmp_key ⊕ cipher_fn
 7:          f_seed ← seed % f_seed_num
 8:          fn_file_name, fn_line_no ← binary_search(f_seed, fn_mapfiles)      ▷ binary_search
 9:          fn_plaintext ← getline(fn_file_name, line_no)[0]               ▷ get firstname plaintext
10:          l_seed ← tmp_key ⊕ cipher_ln
11:          l_seed ← seed % l_seed_num
12:          file_name, line_no ← binary_search(l_seed, ln_mapfile)          ▷ binary_search
13:          ln_plaintext ← getline(fn_file_name, line_no)[0]              ▷ get lastname plaintext
14:          return and response (fn_plaintext,   ln_plaintext)
15:      else
16:          prompt("the username or password is wrong")
17:          return false
18:      end if
19:  end procedure
```

The scheme in this subsection is a comprehensive design that combines designs in previous subsections, and above algorithms are also complex than previous. This scheme provide a practicable SEIGuard solution for server-side information objects which have similar unlimited message space with non-uniform distribution.

### 3.6. SEIGuard System

We develop the SEIGuard system based on the proposed scheme architecture (section 3.1), which integrates all the previous scheme designs and algorithms (section 3.2 ~ 3.5).

| data_1 | data_2 | data_3 | data_4 | data_5 |
|---|---|---|---|---|
| 2281162987101503745 | 2032241654566717917 | 2134976751298369405 | 2281162987101654437 | 1967063960800744414 |
| 8222556215244668687 | 8222556390420902276 | 8222556215240902169 | 8222556215245033011 | 7173546081548367781 |
| 10558774665968163607 | 10558774854893729045 | 10558774183923738968 | 10558774665968285748 | 10964054623074750227 |
| 5843677662448298465 | 5843677441654070573 | 5843676921390272949 | 5843677662448156610 | 6510359720178418747 |
| 7320079739798452959 | 7320079185851516800 | 7320078397717749912 | 7320079739798573595 | 6981262454289139928 |
| 931813156144919952 | 931813649555561947 | 931812636995260786 | 931813156144025155 | 892848586175929297 |

Figure 4: Ciphertexts of social engineering information stored in the database

Figure 5 shows the registration interface of SEIGuard, by which the new user creates an account (username), sets the password and registers their information. All the social engineering information (mobile phone number, identification number, email address and personal name) are gathered together, and the registration of these information objects is completed at one time. The password is processed as a temporary key and dropped after generating the ciphertexts. Thus, only the username and four types of ciphertext are stored in the database. Moreover, these four types of ciphertexts are indistinguishable, since they are honey-encrypted by the same temporary key and have the same ciphertext space (Figure 4). Even if the database is leaked (or compromised), it is just a pile of meaningless data.

Figure 6 shows the interface and responses of current / conventional password-based server system. Sub-figure (a) is the login interface. When a user attempt to log in, the system will check whether the user's password is correct by comparing the password information stored in the database. Sub-figure (b) is the response for a user login (username: wzg018) with the correct password, on which user's plaintext information is returned. Sub-figure (c) shows the response for a user login using an incorrect password (attack), on which the red prompting message (a brute-force anchor) "the username or password is wrong " is returned.

Figure 5: The registration interface

(a)

(b)

(c)

Figure 6: Current / conventional password-based server system. (a) Login interface. (b) Response for correct password login. (c) Response for incorrect password login.

For SEIGuard system, the login interface is the same as Figure 6 (a). However, the procedures behind the login interface is different. SEIGuard doesn't check the validity of the submitted password. As long as the username is valid, no matter the user's password is correct or not, the server directly processes the password to a temporary key and use it to decrypt ciphertexts that stored in the database. When a user (username: wzg018) log in the system with the correct password, Figure 7 (a) shows the response, which is the same as Figure 6 (a). When the user (username: wzg018) log in the system with an incorrect password, after the same backend procedures, SEIGuard system responses (honey plaintexts) as Figure 7 (b) shows. It looks as normal as the Figure 7 (a).

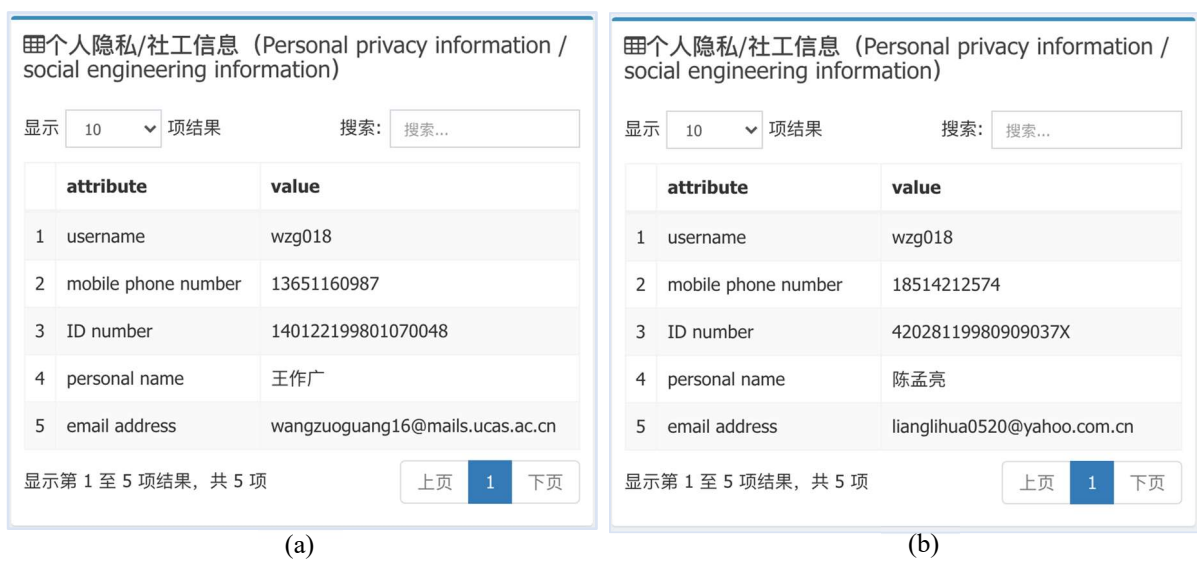

<div align="center">(a)　　　　　　　　　　　　(b)</div>

<div align="center">Figure 7 Response for (a) correct password login vs. (b) incorrect password login</div>

## 4. Experiment and Result

This section evaluates the security and real-time performance of SEIGuard system by two contrast experiments.

### 4.1. Brute-force attack experiment and result

#### 4.1.1. *Attack scenario and defense experiment*

There are many types of brute force attacks. The most representative one is dictionary-based cracking, which employs password dictionaries, hash tables, rainbow tables, etc. If the dictionary covers the entire password space, the dictionary-based cracking becomes an exhaustive brute-force attack. For various types of brute-force attacks, there is a shared core part, that is, they use trial-and-error method to determine whether the guessed password is correct. Therefore, there is no essential difference between online cracking and offline cracking. For a specific password dictionary, no matter what type of the attack is, the result / success rate of the brute-force attack is the same. Therefore, the selected attack scenario is dictionary-based brute-force attack towards the online serve.

The control group uses the classic PBE server system, and the experimental group uses the SEIGuard server system. The two systems run on the same host (the host configuration is shown in later Table 3).

Table 1: attacker's tool and resource

| OS | CPU | RAM | Crack tool | Password dictionary: total lines Username dictionary: total lines | | |
|---|---|---|---|---|---|---|
| VM kali-5.2.0 | Intel(R) Core(TM) i5-1035G1 @1.00GHz | 2 GB | hydra-9.0 | D1: 1000 U1: 30 | D2: 1500 U2: 40 | D3: 2000 U3: 30 |

Table 1 shows the attacker's tool and resource. The attacker employs Hydra [22] as the brute-force attack tool. Three different password dictionaries (D1, D2, D3) are designed based on existing password dictionaries [23][24]. In order to simulate the brute-force cracking using large-volume dictionaries meanwhile shorten the experiment time, these three designed dictionaries don't have a large volume of passwords, but they contain more valid passwords than common dictionaries. The three password dictionaries respectively have 1000, 1500 and 2000 passwords and the corresponding username dictionaries respectively contains 30, 40 and 30 valid usernames.

### 4.1.2. *Experiment result analysis*

Table 2 shows the contrast experiment result of brute-force attack towards conventional PBE scheme and SEIGuard scheme. For each scheme system, three groups of attack experiments were conducted using username dictionary U(1,2,3) and password dictionary D(1,2,3).

In order to evaluate the security of the SEIGuard scheme, we set up three indicators, namely extraction number/rate, verification number/rate (i.e. crack success rate), and average crack success rate. Extraction number/rate, from the third party perspective, shows how many correct password the brute-force attack guessed. Verification number/rate indicates how many the guessed password is (can be) verified to be correct/valid.

In the brute-force attack towards the PBE scheme, in order to speed up the cracking, hydra was set to run 64 crack tasks in parallel. In total, the three control group experiments take about 145000 crack tries which take 76 minutes. The experiment result shows that the extraction rate and verification/success rate of the three dictionaries is respectively 60%, 47.5% and 40%, and the average crack success rate is 49.167%.

Table 2: Experiment result of brute-force attack on the conventional PBE scheme and SEIGuard scheme

| Server | Dictionaries | Users to crack | Crack tries | Time | Extraction num / rate | Verification num / rate | Average crack rate |
|---|---|---|---|---|---|---|---|
| conventional PBE scheme | D1 | 30 | 29741 | 14 min | 18 / 60% | 18 / 60% | |
| | D2 | 40 | 59711 | 32 min | 19 / 47.5% | 19 / 47.5% | 49.167% |
| | D3 | 30 | 55299 | 30 min | 12 / 40% | 12 / 40% | |
| SEIGuard scheme | D1 | 30 | 30 | < 1min | 1 / 29 fake | 0 | |
| | D2 | 40 | 40 | < 1min | 40 fake | 0 | 0 |
| | D3 | 30 | 30 | < 1min | 30 fake | 0 | |

During the brute-force attack towards the SEIGuard scheme, due to the of trial-and-error anchor had been removed, when the first password was tried, even if it was wrong, there was no prompting message showing the password was wrong, nor did the login interface jump to other interfaces except for the normal information page (Figure 7). Figure 7 (a) and (b) are the same interface and share the same server resource such as URL and backend procedures. Thereupon, hydra judged that the first guessing gets the correct password and start the brute-force cracking on the next username. Finally, based on the same dictionaries, the attacker successfully "cracked" all the usernames' passwords in less than three minutes.

But in fact, almost all the passwords cracked by hydra are wrong, except for the first dictionary attack (U1, D1) in experimental group. This is because the first password dictionary is designed based on a weak password dictionary. The password in the first line of D1 happen to be the same password set by one of the 30 user in U1. Consequently, the attacker luckily extracted a correct password.

Successfully cracked all accounts in a short period of time may arouse the attacker's suspicion, since this is very different from the familiar brute-force attack. The attacker may recheck the result take further manual analysis to test new passwords. However, since the SEIGuard system responds correct information (both the content of the information and its occurrence frequency are reasonable) to every password guessing (hydra and manual test), every different guessing will bring about a new "correct" password. And the attacker cannot determine the which one of these "correct" passwords is really valid.

### 4.2. Real-time response experiment and result

This subsection evaluates the real-time response performance of the SEIGuard scheme.

We conducted real-time response comparison experiments between the SEIGuard scheme and the conventional PBE scheme on two different test environments (platform configuration) respectively. The control group server employs the conventional PBE scheme, and the experimental group server applies the SEIGuard scheme. Moreover, in order to make the conventional PBE scheme more representative and reflect the shortest response time, the password was processed by salted hash (default function of *flask_bcrypt, generate_ /check_ password_hash* in Flask 1.1.2). Other user information is stored with plaintext in the database, for which only write and read operations are performed on them in the registration and login responses, and there is no time consumption for encryption and decryption.

To detail the real-time performance, we not only measured the overall response time of registration / login, but also measured the response time of sub-procedures in the registration / login procedure, including password processing, mobile phone number, Chinese ID number, email address, Chinese lastname and Chinese firstname. Moreover, considering the average response time is more representative, we respectively performed five registration tests and five login tests for the control group and the experimental group. Finally, eight average times (in 4 groups) were yielded to analyze the real-time response performance.

Table 3: Test environment 1 (Server with SSD)

| Server OS / Platform | Database Server | CPU | RAM | SSD (Solid State Disk) Information |
|---|---|---|---|---|
| windows 10 64-bit | MySQL 8.0.23 | Intel(R) Core(TM) i5-1035G1 @ 1.00GHz | 16 GB | Disk Random 16.0 Read: 56.10 MB/s Disk Random 16.0 Write: 64.14 MB/s |

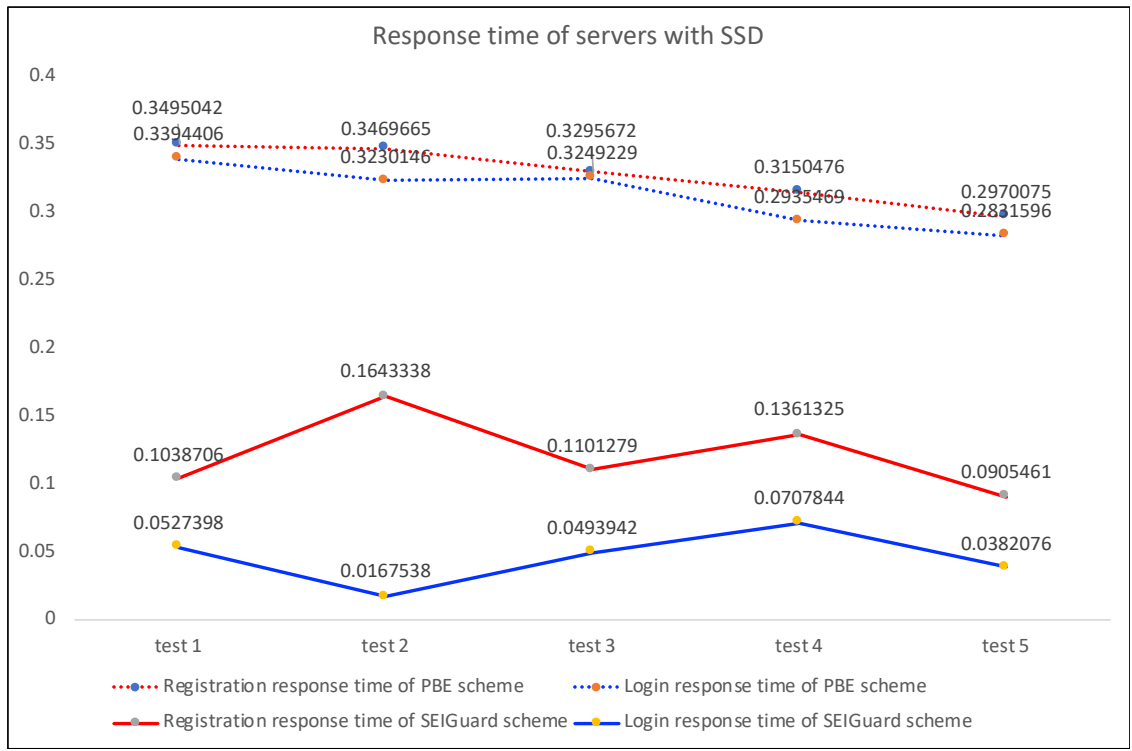

Figure 8: Overall response time of registration / login of servers with SSD

Table 3 shows the configuration of the first test environment (platform), in which the two server systems are hosted on solid state disk (SSD). Figure 8 shows the response time of registration / login of conventional PBE server and SEIGuard server on this platform. It can be seen that the two response timelines of the SEIGuard server are respectively below the response timeline of the conventional PBE server. Table 4 shows the average results of 5 registration/login response time, and also does the statistics of average response time of the sub-procedures. The average registration response time of the conventional PBE scheme is 0.32761860s, which is longer than the average registration response time of the SEIGuard scheme, 0.12101644s. The average login response time of the conventional PBE scheme is 0.31281692s, which is longer than the average login response time of the SEIGuard scheme of 0.03918298s.

Table 4: Real time performance of server with SSD / Average response time of 5 test group

| Information Item | Response time of conventional system | | Response time of SEIGard system | |
|---|---|---|---|---|
| | Registration | Login | Registration | Login |
| password salting hash (PBE) | 0.29735976 | 0.29761686 | - | - |
| mobile phone number | | | 0.00035600 | 0.00005444 |
| identification card number | | | 0.00020562 | 0.00004170 |
| email address | 0.03025884 | 0.01520006 | 0.02395096 | 0.00252718 |
| chinese surname / last name | | | 0.00228522 | 0.00233260 |
| chinese first name | | | 0.07118766 | 0.02838568 |
| Total Response Time | 0.32761860 | 0.31281692 | 0.12101644 | 0.03918298 |

Table 5: Test environment 2 (Server with HDD)

| OS / Platform | Database Server | CPU | RAM | HDD (Hard Disk Drive) Information |
|---|---|---|---|---|
| windows 10 64-bit | MySQL 8.0.12 | Intel(R) Core(TM) i7-4790 @ 3.60GHz | 12 GB | Disk Random 16.0 Read: 1.58 MB/s Disk Random 16.0 Write: 3.27 MB/s |

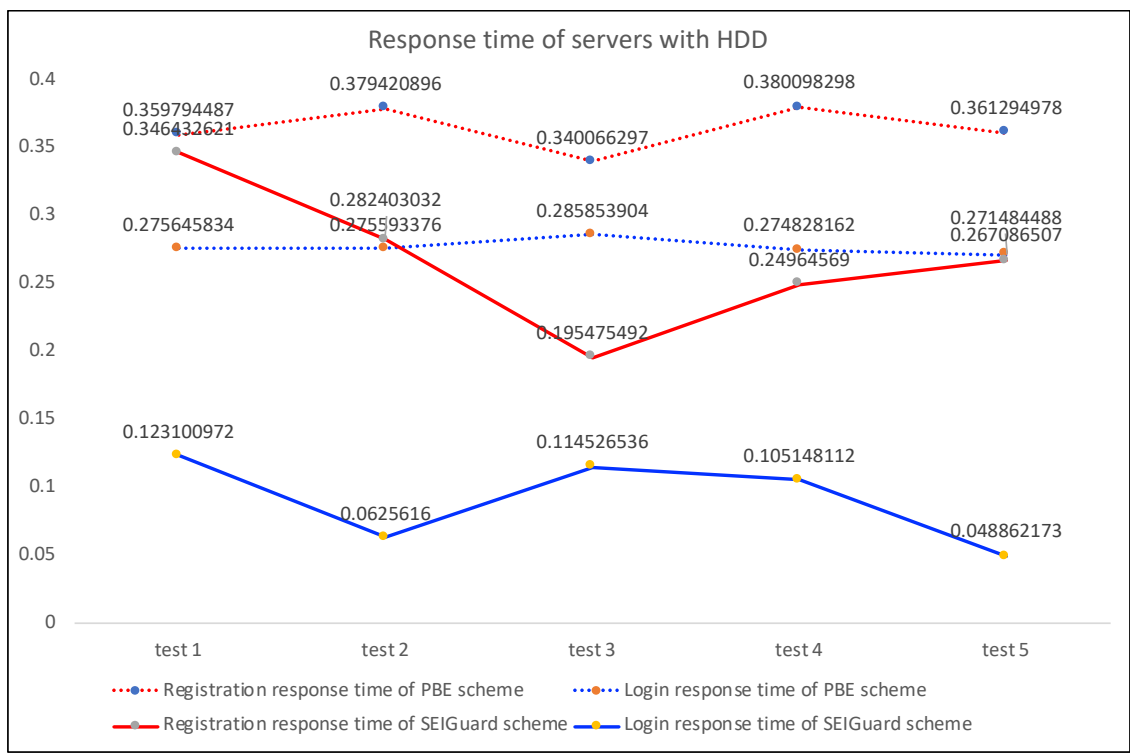

Figure 9: Overall response time of registration / login of servers with HDD

Table 5 shows the configuration of the second test environment (platform). The two server systems are hosted on hard-disk drive (HDD). Figure 9 shows the response time of registration / login of conventional PBE server and SEIGuard server on this platform. Because the random write / read speed of HDD in this test environment is much lower than the speed of SSD, the response time of SEIGuard server has increased slightly. However, similar to Figure 8, the two response timelines of the SEIGuard server are still respectively below the response timeline of the conventional PBE server.

Table 6 shows the average results of 5 registration/login response time, and also includes the average response time of the sub-procedures. In this HDD test environment, the average registration response time of the conventional PBE scheme is 0. 36413499s, which is longer than the average registration response time of the SEIGuard scheme, 0. 26956951s. The average login response time of the conventional PBE scheme is 0. 27668115s, which is longer than the average login response time of the SEIGuard scheme of 0. 09228565s.

Table 6: Real time performance of server with HDD / Average response time of 5 test group

| Information Item | Response time of conventional system | | Response time of SEIGard system | |
|---|---|---|---|---|
| | Registration | Login | Registration | Login |
| password salting hash (PBE) | 0.26666159 | 0.27482406 | - | - |
| mobile phone number | 0.09747340 | 0.00185710 | 0.00154411 | 0.01573814 |

| | | | 0.00188552 | 0.01894036 |
|---|---|---|---|---|
| identification card number | | | 0.00188552 | 0.01894036 |
| email address | | | 0.02982664 | 0.00225192 |
| chinese surname / last name | | | 0.00228614 | 0.00203957 |
| chinese first name | | | 0.10111853 | 0.04538940 |
| Total Response Time | 0.36413499 | 0.27668115 | 0.26956951 | 0.09228565 |

The experiment results under these two different test environments all show that the SEIGuard scheme has a better real-time response performance (a shorter response time) than the conventional PBE scheme.

In addition, it can be said that the configuration of these two test environments is very common, or low. Although the SEIGuard scheme involves numerous read and write operations, encryption and decryption operations, the SEIGuard scheme has a good real-time response performance. Even, the decryption and decryption performance of the larger message space in this paper (phone No.: 0.3~2ms, 0.1~16ms; ID No.: 0.2~2ms, 0.1~19ms,) are superior to the performance in [10] (8 digits password space: max 70s, 180s). This mainly owes to two reasons. (1) Our proposed strategy of multiple small mapping files and the designed binary search algorithms for all information location/search, the time complexity of which is only O(log n). (2) The salted hash for password in the conventional PBE scheme takes a little time, but the SEIGuard scheme simplifies the authentication and doesn't require password hashing and storage operations.

Overall, the real-time response performance of the SEIGuard scheme is impressive. Even under conditions of very ordinary hardware configuration, the registration response (0.121s~0.269s) and login (0.039s~0.092s) responses still work smoothly.

## 5. Discussion

About the applicability and practicality of the SEIGuard scheme. As the applicability analysis mentioned at the end of section 3.2~3.5 and the real-time performance that experiments shows, SEIGuard scheme is also applicable and practical to protect other social engineering information whose message space is similar to the four types message space in section 3. As a summary, SEIGuard scheme can be used to protect social engineering information such as mobile phone number, (Chinese) identification number, email address, (Chinese) personal name, personal password, pin code, bank debit / credit card number and password, social security number, IP address, occupation & job and title information, and location information.

About the user information / privacy leakage. For SEIGuard scheme, a brute force attacker can easily obtain a lot of information, since for every guessing the server responds with a message from the message space. These messages may correspond to (or contain) other users' information/privacy. However, this is not a problem. Because the attacker can still obtain / generate these messages independently, such as numerous mobile phone numbers and ID numbers, without interacting with the SEIGuard system. For SEIGuard scheme on email addresses, brute-force attackers may get a lot of real and honey email addresses. However, the these real and honey message is indistinguishable for attackers. If the attacker merely use the honey email address to learn the target's affiliation by the domain-part information, the attacker will be misled; if the attacker use the guessed email address to launch a phishing attack, due to the mismatched personal information, the phishing attack is relatively easy to perceive. Besides, other defense measures can be combined to enhance the security. For instance, the honey email address can be monitored by security administrator / detect system, and once an abnormal status was detected, it triggers an alarm. In addition, limiting the logins number within a certain period of time is also a mitigation measure. For SEIGuard scheme on personal name, the returned/revealed honey personal names to attackers is like a name that a child randomly fabricates / lies from name space (based on the message probability distribution). Taken as the whole responses, it will not reveal the privacy of other users.

About the problems caused by typos. When legitimate users mistakenly type a wrong password in the login interface, they also get the fake plaintexts that look real. This is an open problem of honey encryption (HE) [12]. The deception response to decryption using incorrect key (no matter it is intentionally or unintentionally) is an inherent requirement of HE to achieve security. For SEIGuard scheme, typos is a limitation, meanwhile may be a merit. It is precisely the indiscriminately treating to legal users (type a correct password) and illegal users (type a correct password) that eliminates the security risk of brute-force attacks. We tried to deal with this problem, but the remediation for typos of SEIGuard may be a complex problem. Because while we provide typos solution to legitimate users, we also provide the same help to attackers. And we will study this problem in future works.

# 6. Conclusions

This paper proposes the SEIGuard: an authentication-simplified and deceptive scheme to protect server-side social engineering information against brute-force attacks. In the SEIGuard scheme, the password check procedure is omitted and there is no password or encrypted/processed password stored in the database. And this design is further combined with the (users' information) data encryption design using honey encryption / decryption, in which the login password merely serves as a temporary key to generate / decrypt the ciphertexts and is later dropped. The login using correct password will decrypt the ciphertexts to correct (real / original) plaintext. On the contrary, the login using incorrect password will decrypt the ciphertext to a honey / fake but plausible-looking plaintext, which is sampled / mapped form the same message space (and subject to the same message distribution) as the correct plaintext belongs. The honey plaintext obtained by incorrect login password will mislead attackers. No matter the passwords submitted in the login interface are correct or wrong, they share the same undifferentiated backend procedures. This scheme eliminates the anchor that both the online and offline brute-force attacks depending on. Furthermore, besides the SEIGuard scheme architecture, this paper proposes four SEIGuard scheme designs and algorithms for 4 typical social engineering information objects (mobile phone number, identification number, email address, personal name), which represent 4 different types of message space, i.e. 1) limited and uniformly distributed, 2) limited, complex and uniformly distributed, 3) unlimited and uniformly distributed, 4) unlimited & non-uniformly distributed message space. Specially, we propose multiple small mapping files strategies, binary search algorithms, two-part HE (DTE) design and incremental mapping files solutions for the applications of SEIGuard scheme on information objects belongs to the above 4 message space. Finally, this paper develops the SEIGuard system based on the proposed schemes, designs and algorithms. Contrast experiments show that the SEIGuard scheme can effectively protect server-side social engineering information against brute force attacks. The experiment evaluation also shows that despite the very ordinary test environment, SEIGuard scheme has an impressive real-time response performance that is better than conventional PBE server scheme and HE encryption and decryption.